\documentstyle[aps,multicol,amsmath]{revtex}
\tighten
\newcommand{\bleq}{\ifpreprintsty \else
\end{multicols}\vspace*{-3.5ex}{\tiny \noindent\begin{tabular}[t]{c|}
\parbox{0.493\hsize}{~} \\ \hline \end{tabular}} \fi}
\newcommand{\eleq}{\ifpreprintsty \else
{\tiny\hspace*{\fill}\begin{tabular}[t]{|c}\hline
\parbox{0.49\hsize}{~} \\
\end{tabular}}\vspace*{-2.5ex}\begin{multicols}{2} \fi}
\newcommand{\bcols}{\ifpreprintsty\else\begin{multicols}{2}\fi}
\newcommand{\ecols}{\ifpreprintsty\else\end{multicols}\fi}
\newcommand{\non}{}
\newcommand{\be}{\begin{eqnset}\,}
\newcommand{\ee}[1]{#1\end{eqnset}}

\newcommand{\nn}{\global\let\non=\nonumber}

\newcommand{\sfrac}[2]{\mbox{$\frac{#1}{#2}$}}

\newcommand{\saverage}[1]{\langle#1\rangle}
\newcommand{\average}[1]{\left\langle#1\right\rangle}

\newcommand{\eql}[1]{\global\let\non=\relax\label{eq:#1}%
\quad^{\underline{\{#1\}}}
}
\newcommand{\eq}[1]{(\ref{eq:#1})}

\newlength{\spacelen}
\newenvironment{eqnset}
{%
 \global\let\non=\nonumber
  \vspace{-0.6ex}
 \begin{eqnarray}%
 \everymath{\displaystyle}%
 \begin{array}{[r@{\;\:}c@{\;\:}l]@{\vspace{1.6ex}}}%
}%
{%
 \vspace{-1.0ex}%
 \end{array}\non\end{eqnarray}\hspace{-\spacelen}%
}
\newlength{\ddone}\ddone 0pt
\newlength{\lzero}\lzero 0pt
\newcommand{\dd}[2]{%
        \settowidth{\ddone}{$#1$}%
        \ifdim\ddone=\lzero%
                \partial_{#2}%
        \else%
                {\frac{\partial #1}{\partial #2}}%
        \fi%
}
\newlength{\ddhone}\ddone 0pt
\newcommand{\ddh}[2]{%
        \settowidth{\ddhone}{$#1$}%
        \ifdim\ddhone=\lzero%
                (d/d#2)%
        \else%
                {\frac{d #1}{d #2}}%
        \fi%
}

\newcommand{\ccontract}{{\,{:}\,}}
     
\renewcommand{\vec}[1]{{\bf #1}}
\renewcommand{\eql}[1]{\label{eq:#1}}
\renewcommand{\eq}[1]{Eq.~(\ref{eq:#1})}
\newcommand{\baverage}[1]{\bigl\langle #1 \bigr\rangle}

\newcommand{\barM}{{\bar M}}

\newcommand{\calP}{{\cal P}}

\newcommand{\calL}{{\cal L}}
\let\oldperp\perp
\renewcommand{\perp}{{\!\oldperp}}
 
\begin{document}
 
\title{Multiple-Point and Multiple-Time Correlations Functions in a
Hard-Sphere Fluid }
\author{Ramses\ van Zon and Jeremy\ Schofield}
\address{Chemical Physics Theory Group, Department of Chemistry,\\
University of Toronto,Toronto, Ontario, Canada M5S 3H6}
\date{\today}
\maketitle
 
\begin{abstract} 
A recent mode coupling theory of higher-order correlation functions
is tested on a simple hard-sphere fluid system at intermediate densities.
Multi-point and multi-time correlation functions of the densities of
conserved variables are calculated in the hydrodynamic limit and
compared to results obtained from event-based molecular dynamics
simulations.  It is demonstrated that the mode coupling theory
results are in excellent agreement with the simulation results
provided that dissipative couplings are included in the vertices
appearing in the theory.  In contrast, simplified mode coupling
theories in which the densities obey Gaussian statistics neglect
important contributions to both the multi-point and multi-time
correlation functions on all time scales.   
\vspace{1em} 
\end{abstract}
\bcols

\section{Introduction}

Over the last few years, the emergence of multi-dimensional
NMR\cite{Sillescu96,Spiess98} and non-resonant non-linear Raman\cite{Fleming,Miller,Tokmakoff} 
techniques has generated renewed interest in the information content of
higher-order correlation functions involving time-correlations of dynamical
quantities at multiple points and time separations.  These
experimental developments hold great promise for the elucidation of the
nature of the underlying dynamics giving rise to complex relaxation
behavior in super-cooled liquids, polymeric systems, and
proteins\cite{Mukamel99,Steffen97,Keyes2000}.  
Concurrently, simulation
studies probing the microscopic origin of dynamical heterogeneity in
dense systems\cite{Kob} have made use of the increased information
content available in multiple-point\cite{Glotzer} and
multiple-time\cite{Heueretal} correlation functions.  

Although there
has been some recent work attempting to reproduce simulation results
for the off-resonant fifth-order Raman response function\cite{Tanimura97,Stratt,Reichman}, 
there has been little theoretical work to establish a
microscopic theory for general higher-order correlation functions.  In
a previous article\cite{vanZon}, a general mode coupling theory was
developed in which the long time behavior for multi-point and
multi-time correlation functions was expressed in terms of ordinary
two-time, two-point correlation functions of a set of slow
variables which are coupled by vertices containing both static (called Euler) and
dynamic (called dissipative) correlations.  The theory is based upon
the assumption that the long-time dynamics of arbitrary variables is a
functional of a set of slow modes of the system.  
The long-time dynamics of higher-order correlation functions is then
described by isolating the component of the relevant variables along
multi-linear products of the slow variables, resulting in expressions
for the higher-order correlation functions in terms of the sum of an
infinite number of multi-point correlation functions of slow modes.
The formulation is
made tractable by a cumulant expansion method (called
$N$-ordering\cite{MO82,SLO92})
in which multi-point
correlation functions are factored into convolutions over the familiar
two-point, two-time correlation functions of the
slow modes.  In this way, the need to simplify the
mode-coupling expressions for higher-order correlation functions based
on an assumption of Gaussian statistical behavior of the slow modes is
avoided.  It was suggested that simple mode coupling theories\cite{K70,R81} based
upon this Gaussian assumption lead to a relatively poor description of
the long time behavior of higher-order correlation functions.

The purpose of this article is to validate the mode coupling theory
expressions for multi-point and multi-time correlation functions by examining the
simplest non-trivial system, the hard-sphere liquid.  The hard-sphere
liquid is a very useful system to examine theoretically since the
simple form of the interaction potential allows static correlation
functions to be related to the radial distribution function at
contact.  In turn, the radial distribution function can be
approximated using an accurate equation of state, such as that of
Carnahan and Starling\cite{Carnahan}, which relates the pressure to
the density and the temperature.  In addition, excellent
predictions exist for dynamical properties of hard sphere systems based
on detailed kinetic theory\cite{Chapman}.  Another advantage of
looking at hard sphere systems is that 
the dynamics of the system can be simulated very efficiently using
event-based molecular dynamics methods\cite{eventMD} since particles
evolve freely between collisions, thereby allowing good statistics
to be obtained from simulations.  

We shall focus on systems of moderate reduced densities ($\rho^{*} =0.25$) in which
``mode coupling'' effects leading to non-exponential relaxation of
correlation functions of linear densities, such as the dynamical
structure factor, can be neglected. In particular, we target
correlation functions of long-wavelength fluctuations which decay on long time scales
and exhibit complicated higher-order correlation functions.  

This paper is organized as follows:  In Section II, the mode coupling
formalism developed in Ref.~\cite{vanZon} is reviewed and adapted to
the hard sphere system.  Explicit expressions are presented for 
three-point and three-time correlation
functions of involving linear densities of number (or mass), transverse, and
longitudinal velocities.  In Section III, simulation methods particularly
suited for calculating higher-order correlation functions in a hard-sphere system
are discussed.  In Section IV, the predictions of
the mode coupling theory are compared to the simulation results for
relatively simple three-point and three-time correlation functions,
and it is demonstrated that dissipative parts of vertices provide additional important
couplings to those at Euler order.  The results are contrasted
with those obtained within the framework of Gaussian mode coupling
theory\cite{K70,R81}.  Finally, conclusions of the study are given in
Section V.

\section{Theoretical formulation}

The system under consideration is composed of $N$ particles of
mass $m$ and diameter $a$ in a volume $V=L_x \times L_y \times L_z$.
The particles interact through the two-body hard-sphere potential,
\begin{equation}
V(r)=\left\{\begin{array}{ll}
                0       & \mbox{if $r < a$}\\      
		\infty & \mbox{if $r \leq a$} .
\end{array} \right.
\end{equation}
Given the form of the potential, the dynamics generated by the
Hamiltonian conserves the total number of particles $N$, the total
angular momentum, the linear momenta $\vec{P}$, and the energy $E$ of the system.
In Ref.~\cite{vanZon}, expressions for the long time behavior of correlation functions
were obtained under the assumption that the slowly varying part of an
arbitrary dynamical variable is an analytic function of a set of slow
variables $\vec{A}$ of the system.  An essential part of 
successfully applying the formalism to a particular system
is the identification of a {\it complete} set of slow variables.  To
identify the slow modes of the system, it is helpful to consider the
local densities of the conserved variables $N$, $\vec{P}$ and $E$,
\begin{eqnarray*}
N(\vec{r}) &=& \sum_{i=1}^{N} \delta (\vec{r}-\vec{r}_{i}) , \\
\vec{P}(\vec{r}) &=& \sum_{i=1}^{N} \vec{p}_{i} \delta
(\vec{r}-\vec{r}_{i}) , \\
E(\vec{r}) &=& \sum_{i=1}^{N} \left( \frac{p_{i}^{2}}{2m}+ \frac{1}{2}
\sum_{j \neq i} V(|\vec{r}_{i}-\vec{r}_{j}|) \right)\delta (\vec{r}-\vec{r}_{i}) ,
\end{eqnarray*}
where $\vec{r}_{i}$ and $\vec{p}_{i}$ are the spatial position and
momentum of particle $i$.
Noting that the Fourier transform of these densities,
\begin{eqnarray}
N_{\vec{k}} &=& \sum_{i=1}^{N} e^{i\vec{k}\cdot \vec{r}_{i}},
\nonumber \\
\vec{P}_{\vec{k}} &=& \sum_{i=1}^{N} \vec{p}_{i} e^{i\vec{k}\cdot \vec{r}_{i}},
\label{linearDensities}\\
E_{\vec{k}} &=& \sum_{i=1}^{N} \left( \frac{p_{i}^{2}}{2m} +
\frac{1}{2} \sum_{j \neq i} V(|\vec{r}_{i}-\vec{r}_{j}|) \right)e^{i\vec{k}\cdot \vec{r}_{i}},
\nonumber
\end{eqnarray}
are slowly varying quantities for small $k=|\vec{k}|$ since their time
derivatives are proportional to $k$, the minimal set of slow variables
$A_{\vec{k}}'$
must include all the ``hydrodynamic'' variables $\{ N_{\vec{k}},
\vec{P}_{\vec{k}}, E_{\vec{k}} \}$ with $k$ smaller than some cut-off
wave vector $k_{c}$.  For our purposes, it is
convenient to work with a slightly different basis set $A_{\vec{k}}$, composed of 
the variables $N_{\vec{k}}$, $L_{\vec{k}}=\vec{P}^{x}_{\vec{k}}$, $T_{1\vec{k}}=\vec{P}^{y}_{\vec{k}}$,
$T_{2\vec{k}}=\vec{P}^{z}_{\vec{k}}$, and $H_{\vec{k}}=(3 N_{\vec{k}}-2\beta
E_{\vec{k}})/\sqrt{6}$, where $\beta=1/(k_{B}T)$ is the inverse
temperature of the system, $\vec{P}_{\vec{k}}^{x}$ is the
$x$-component of the vector $\vec{P}_{\vec{k}}$, and $\hat{\vec{k}}$ is taken along the
$x$-axis.  Note that the $T_{1\vec{k}}$ and $T_{2\vec{k}}$ are the transverse modes of
the momentum density, while $L_{\vec{k}}$ is the longitudinal momentum
density.  With this definition of basis set, the matrix
\begin{eqnarray*}
\saverage{A^{a}_{\vec{k}}A^{b*}_{\vec{k}}}
=\saverage{A^{a}_{\vec{k}}A^{b}_{-\vec{k}}} \delta_{a,b}
\end{eqnarray*}
is diagonal in the hydrodynamic labels $a$ and $b$, 
where $\saverage{\cdots}$ denotes the grand-canonical ensemble average.
The non-linear dependence of the dynamical variables is expressed in
terms of a ``multi-linear'' basis set
        \begin{eqnarray}	
                Q_0 &\equiv& 1 
        \nonumber\\
                Q_1 &\equiv& A_{\vec{k}} - \average{A_{\vec{k}}}
\equiv \hat{A}_{\vec{k}}
        \eql{Qset}\\
                Q_2 &\equiv& Q_{\vec{k}-\vec{q}}Q_{\vec{q}}
                -\average{Q_{\vec{k}-\vec{q}}Q_{\vec q}}
                -\average{Q_{\vec{k}-\vec{q}}Q_{\vec{q}}Q_1^*}\cdot
K_{11}^{-1} \cdot Q_1, 
         \nonumber\\
         \vdots \nonumber 
        \end{eqnarray}
where the ``$\cdot$'' notation denotes a sum over components of the column
vector $A_{\vec{k}}$ (the indices of the hydrodynamic variables
$N_{\vec{k}}$, $L_{\vec{k}}$, $T_{1\vec{k}}$, $T_{2\vec{k}}$, and $H_{\vec{k}}$).
The subtractions in the basis set defined in \eq{Qset} are included to ensure that the
multi-linear matrix, 
\begin{equation}
K_{lm}=\saverage{Q_{l} Q_{m}^{*}}=\saverage{Q_{l} Q_{m}^{*}} \;
\delta_{l,m} , 
\eql{Kdef}
\end{equation} is diagonal in
{\it mode-order} $l$.  The slow part of any
dynamical variable $C$ is removed by the projection operator
\begin{equation}
        \calP C \equiv \sum_{l=0}^{\infty}\saverage{C Q^*_l}
        K^{-1}_{ll}  Q_{l},
\end{equation} 
and the complementary projection operator ${\cal{P}}_{\perp}=1-\cal{P}$
projects onto the orthogonal sub-space.         

Writing the three-point correlation function 
$\saverage{\hat{A}_{\vec{k-q}}(t) \hat{A}_{\vec{q}}(t)
\hat{A}_{-\vec{k}}}$ in terms of the basis set, we obtain

\begin{eqnarray}
\saverage{\hat{A}_{\vec{k-q}}(t) \hat{A}_{\vec{q}}(t)
\hat{A}_{-\vec{k}}} &=& \saverage{\hat{A}_{\vec{k}}(t)
\hat{A}_{-\vec{k}}}
\cdot K_{11}^{-1} \cdot \saverage{ \hat{A}_{-\vec{k}}
\hat{A}_{\vec{k-q}}\hat{A}_{\vec{q}}} \nonumber \\
&& + G^{21}_{\vec{k-q},\vec{q};\vec{k}}(t) \cdot K_{11},
\eql{generalMP}
\end{eqnarray}
where $G^{mn}(t)=\saverage{Q_{m}(t)Q_{n}^{*}} * K_{nn}^{-1}$, and, in particular,
\begin{eqnarray}
G^{21}_{\vec{k-q},\vec{q};\vec{k}}(t)= \frac{\saverage{Q_{2}(\vec{k-q},\vec{q}; t)
\hat{A}_{-\vec{k}}}}{\saverage{\hat{A}_{\vec{k}} \hat{A}_{-\vec{k}}}}.
\eql{G21def}
\end{eqnarray}
Note that \eq{generalMP} is exact in the limit $t
\rightarrow 0$ by construction of the basis set.  
Utilizing projection operator
techniques\cite{Zwanzig61,Mori65,SLO92} and cumulant expansion
methods\cite{MO82}, the multi-point correlation function
$G^{21}_{\vec{k-q},\vec{q};\vec{k}}(t)$ can be expressed in terms of
two-point, two-time correlation functions as\cite{vanZon},

\begin{eqnarray}
        G^{21}_{\vec{k-q},\vec{q};\vec{k}}(t)
        &=&
        \int_0^t
        G^{11}_{\vec{k-q}}(t-\tau) G^{11}_{\vec{q}}(t-\tau) 
\nonumber \\ &&
        \ccontract
        \barM^{21}_{\vec{k-q},\vec{q};\vec{k}} \cdot
G^{11}_{\vec{k}}(\tau)d\tau ,
\eql{G21}
\end{eqnarray}            
where $G^{11}_{\vec{k}}(t)=\saverage{\hat{A}_{\vec{k}}(t)
\hat{A}_{-\vec{k}}} / \saverage{\hat{A}_{\vec{k}} \hat{A}_{-\vec{k}}}$ are the normalized, two-point
and two-time correlation functions of the linear densities, and the ``vertices'' are given by
\begin{eqnarray}
\barM^{lm} &=& \left[ \saverage{\dot{Q_{l}}Q_{m}^{*}} - 
\int_{0}^{\infty} d\tau \; \saverage{\phi_l(\tau)\phi^*_{m}}\right]
         K^{-1}_{mm},
\eql{vertex}     
\end{eqnarray}
with the fluctuating force $\phi_l(t)$ defined by
\begin{equation}
        \phi_l(t) \equiv
e^{ (1-{\cal P})\calL t } (1-{\cal P}) \dot{Q}_l
\eql{flucforce}
,
\end{equation}                                                    
where $\cal{L}$ is the Liouville operator.

Similarly, it can be shown that the three-time correlation function,
\begin{eqnarray}
G^{111}(t_{1},t_{2})=\saverage{Q_{1}(t_1+t_2)Q_{1}(t_{1})Q_{1}^{*}}
\cdot K_{11}^{-1}
\end{eqnarray}
can be approximately written as\cite{vanZon}, 
\begin{equation}
\begin{split}
        G^{111}(t_2,t_1)=& G^{11}(t_2)
        * \barM^{111} * G^{11}(t_1)
\\+&             G^{12}(t_2) * \barM^{211} * G^{11}(t_1)
\\+&             G^{11}(t_2) * \barM^{112} * G^{21}(t_1)
+ O(N^{-1}) ,
\eql{generalMT}
\end{split}
\end{equation}              
where $\barM^{lmn}$ is given by
\begin{eqnarray}
        \barM^{lmn}&=& 
        \saverage{Q_l Q_m Q_n^{*}} \cdot K^{-1}_{nn}
\eql{M2}.
\end{eqnarray}  
Furthermore, it was shown in Ref.~\cite{vanZon} that $G^{12}(t_{2})$
can be written in terms of the two-point, two-time functions and the
$\barM^{21}$ vertices in a manner analogous to \eq{G21}.  

The symmetry properties of the Hamiltonian can be used to greatly
simplify the analysis of higher-order correlation functions.  For
example, since the Hamiltonian $\cal H$ is invariant under the transformation
${\cal{TH}}=\cal H$, where the self-adjoint time-reversal operator
$\cal{T}$ acts on an arbitrary phase point $(\vec{r}^{N},\vec{p}^{N})$ by
${\cal{T}} (\vec{r}^{N},\vec{p}^{N}) = (\vec{r}^{N}, -\vec{p}^{N})$,
all time correlation functions considered here have well-defined
symmetry properties under $\cal{T}$:
${\cal{T}} \hat{A}^{a}_{\vec{k}}=\gamma_{a} \hat{A}^{a}_{\vec{k}}$,
where $\gamma_{a}=1$ for $a=N,H$ and $\gamma_{a}=-1$ for
$a=T_{1},T_{2}, L$.  Furthermore, since the Liouville operator
$\cal{L}$ transforms as $\cal{T}\cal{L}=-\cal{L} \cal{T}$, it is easy to
show\cite{vanZon,BerneandPecora} that $\saverage{A^{a}_{\vec{k}}(t)
A^{b}_{-\vec{k}}}= \gamma_{a} \gamma_{b} \saverage{A^{a}_{\vec{k}}(-t)
A^{b}_{-\vec{k}}}$.  It is straightforward to extend these arguments
to multi-time correlation functions for which
$\saverage{A^{a}_{\vec{k-q}}(t_1+t_2) A^{b}_{\vec{q}}(t_1) 
A^{c}_{-\vec{k}}}= \gamma_{a} \gamma_{b} \gamma_{c} \saverage{A^{a}_{\vec{k}}(-t_1-t_2)
A^{b}_{\vec{q}}(-t_1)A^{c}_{-\vec{k}}}$.

\subsection{Three-point correlations}

We now turn our attention to evaluating the expressions for
three-point correlation functions of three linear densities of the form  in \eq{generalMP} for several
different combinations of wave-vector and hydrodynamic labels in terms
of the linear-linear correlation functions $G^{11}(t)$.  For
simplicity, we consider correlation functions involving the transverse
momentum mode $T_{2\vec{k}}$ henceforth abbreviated as just
$T_{\vec{k}}$.  From symmetry considerations it is easy to establish
that the linear-linear correlation function $G_{\vec{k}}^{T a}(t)=0$
unless $a=T$, which simplifies the subsequent
analysis.

Looking first at the correlation function,
$\saverage{T_{\vec{k-q}}(t) 
T_{\vec{q}}(t) \hat{N}_{-\vec{k}}}$, using \eq{generalMP} we have,
\begin{eqnarray}
\saverage{T_{\vec{k-q}}(t) T_{\vec{q}}(t)
\hat{N}_{-\vec{k}}} &=& \frac{\saverage{\hat{N}_{\vec{k}}(t)
\hat{A}^{a}_{-\vec{k}}}}
{\saverage{\hat{A}^{a}_{\vec{k}} \hat{A}^{a}_{-\vec{k}}}}  \saverage{ \hat{A}^{a}_{-\vec{k}}
T_{\vec{k-q}} T_{\vec{q}}} \nonumber \\
&& + G_{\vec{k-q},\vec{q};\vec{k}}^{TT;N}(t) \saverage{
\hat{N}_{\vec{k}} \hat{N}_{-\vec{k}}},
\eql{TTN1}
\end{eqnarray}
where the repeated index $a$ is summed over the labels for $N, T, L$
and $H$, and 
\[
G_{\vec{k-q},\vec{q};\vec{k}}^{TT;N}(t) \equiv 
\frac{\saverage{Q_{2}^{TT}(\vec{k-q},\vec{q},t)
\hat{N}_{-\vec{k}}}}{\saverage{\hat{N}_{\vec{k}} \hat{N}_{-\vec{k}}}}.
\]
The replacement of the ``$21$'' super-script in
$G^{21}_{\vec{k-q},\vec{q};\vec{k}}(t)$ by ``$TT;N$'' above is meant to
denote the specific hydrodynamic labels under consideration.  The
semi-colon separating the labels indicates that the labels ``$TT$''
correspond to the bi-linear density, whereas the ``$N$'' labels the linear
density in \eq{G21def}.
Noting that $\saverage{ \hat{A}^{a}_{-\vec{k}}
T_{\vec{k-q}}T_{\vec{q}}}$ vanishes unless $a=N,H$, 
the first part of \eq{TTN1} can be written as,
\begin{equation*}
\begin{split}
\frac{\saverage{\hat{N}_{\vec{k}}(t)
\hat{N}_{-\vec{k}}}}{S(k)} \saverage{ \hat{N}_{-\vec{k}}
T_{\vec{k-q}}T_{\vec{q}}}
+ \frac{\saverage{\hat{N}_{\vec{k}}(t)
\hat{H}_{-\vec{k}}}}{\saverage{N}} \saverage{ \hat{H}_{-\vec{k}}
T_{\vec{k-q}}T_{\vec{q}}} \nonumber
\\
= \frac{m}{\beta} \saverage{\hat{N}_{\vec{k}}(t)
\hat{N}_{-\vec{k}}}-\frac{2m}{\sqrt{6}\beta} \saverage{\hat{N}_{\vec{k}}(t)
\hat{H}_{-\vec{k}}},
\end{split}
\end{equation*}
where $S(k)=\saverage{\hat{N}_{\vec{k}}\hat{N}_{-\vec{k}}}$ is the
static structure factor. The normalized multi-point correlation
function $G_{\vec{k-q},\vec{q};\vec{k}}^{TT;N}(t)$ of the
basis set variable $Q_{2}^{TT}$ can be evaluated using \eq{G21},
\begin{eqnarray}
        G^{TT;N}_{\vec{k-q},\vec{q};\vec{k}}(t)
        &=&
        \int_0^t
        G^{TT}_{\vec{k-q}}(t-\tau) G^{TT}_{\vec{q}}(t-\tau) 
\nonumber \\ &&
        \barM^{TT;a}_{\vec{k-q},\vec{q};\vec{k}} 
G^{aN}_{\vec{k}}(\tau)d\tau ,
\eql{TTN2}
\end{eqnarray}      
where $a$ is summed over the labels $N$ and $H$ only since the
$\barM^{TT;a}$ vertex vanishes when $a=L$ or $T$.  The explicit form
of the vertix is given by \eq{vertex},
which involves an ``Euler part'',
\begin{eqnarray*}
\frac{ \average{\dot{Q}^{TT}_{\vec{k-q},\vec{q}}
\hat{A}^{a}_{-\vec{k}}}}{\saverage{ \hat{A}^{a}_{\vec{k}} \hat{A}^{a}_{-\vec{k}}}}
\end{eqnarray*}
and a dissipative part 
\begin{eqnarray*}
       - \int_{0}^{\infty} d \tau
\saverage{\phi^{TT}_{\vec{k-q},\vec{q}}(\tau) \phi^{a}_{-\vec{k}}} /
\saverage{ \hat{A}^{a}_{\vec{k}} \hat{A}^{a}_{-\vec{k}}} .
\end{eqnarray*}
These contributions to the vertex can be evaluated as detailed in the appendix, and one finds that to
leading order in the wave-vector, only the $a=L$ term contributes
at order $k$ (with corrections of order $k^{3}$),
whereas the other vertex with $a=H$ gives a contribution proportional to
$k^{2}$.  Putting all this together, we obtain the expression,
\begin{align}
\saverage{T_{\vec{k-q}}(t) T_{\vec{q}}(t)
\hat{N}_{-\vec{k}}} &= \frac{m}{\beta} \saverage{\hat{N}_{\vec{k}}(t)
\hat{N}_{-\vec{k}}} -\frac{2m}{\sqrt{6}\beta} \saverage{\hat{N}_{\vec{k}}(t)
\hat{H}_{-\vec{k}}} \nonumber \\
&+ G_{\vec{k-q},\vec{q};\vec{k}}^{TT;N}(t) 
\saverage{\hat{N}_{\vec{k}} \hat{N}_{-\vec{k}}} ,
\end{align}
where
\begin{equation}
\begin{gathered}
G_{\vec{k-q},\vec{q};\vec{k}}^{TT;N}(t) = \label{Gttn}\\
  \int_0^t
        G^{TT}_{\vec{k-q}}(t-\tau) G^{TT}_{\vec{q}}(t-\tau) 
        \barM^{TT;L}_{\vec{k-q},\vec{q};\vec{k}} 
G^{LN}_{\vec{k}}(\tau)d\tau  \\
+ \int_0^t
        G^{TT}_{\vec{k-q}}(t-\tau) G^{TT}_{\vec{q}}(t-\tau) 
        \barM^{TT;H}_{\vec{k-q},\vec{q};\vec{k}} 
G^{HN}_{\vec{k}}(\tau)d\tau ,
\end{gathered}
\end{equation}
and the functions $G_{\vec{k}}^{TT}(\tau )$, $G_\vec{k}^{LN}(\tau )$, and $G_{\vec{k}}^{HN}(\tau
)$ are given explicitly by
\begin{eqnarray*}
G_{\vec{k}}^{TT}(\tau ) &=& \saverage{ T_{\vec{k}}(\tau)
T_{-\vec{k}}}/ m \saverage{N} k_{B}T \\
G_{\vec{k}}^{LN}(\tau ) &=& \saverage{ L_{\vec{k}}(\tau)
\hat{N}_{-\vec{k}}}/ S(k) \\
G_{\vec{k}}^{HN}(\tau ) &=& \saverage{ H_{\vec{k}}(\tau)
\hat{N}_{-\vec{k}}}/ S(k).
\end{eqnarray*}

The vertices $\barM^{TT;L}_{\vec{k-q},\vec{q};\vec{k}}$ and 
$\barM^{TT;H}_{\vec{k-q},\vec{q};\vec{k}} $ are given in the appendix.
Note that if the dissipative parts of the vertices are neglected, only the first
time convolution integral in Eq.~(\ref{Gttn}) contributes to 
$\saverage{T_{\vec{k-q}}(t) T_{\vec{q}}(t)
\hat{N}_{-\vec{k}}}$.

From similar considerations, it is not difficult to
obtain expressions for other correlation functions.  For example, we
find the multi-point functions $\saverage{T_{\vec{k-q}}(t) L_{\vec{q}}(t)
T_{-\vec{k}}}$ and $\saverage{T_{\vec{k-q}}(t) \hat{N}_{\vec{q}}(t)
T_{-\vec{k}}}$ are given by
\begin{eqnarray}
\saverage{T_{\vec{k-q}}(t) L_{\vec{q}}(t)
T_{-\vec{k}}} &=& G_{\vec{k-q},\vec{q};\vec{k}}^{TL;T}(t)
\label{TLTfinal}\\
\saverage{T_{\vec{k-q}}(t) \hat{N}_{\vec{q}}(t)
T_{-\vec{k}}} &=& \frac{S(q)}{\average{N}}
\saverage{T_{\vec{k}}(t)T_{-\vec{k}}} + 
G_{\vec{k-q},\vec{q};\vec{k}}^{TN;T}(t) \label{TNTfinal}
\end{eqnarray}
with
\begin{eqnarray}
G_{\vec{k-q},\vec{q};\vec{k}}^{TL;T}(t) &=& \int_0^t
G^{TT}_{\vec{k-q}}(t-\tau) \label{Gtlt} \\
 && \sum_{a=L,N,H} G^{La}_{\vec{q}}(t-\tau) 
        \barM^{Ta;T}_{\vec{k-q},\vec{q};\vec{k}} 
G^{TT}_{\vec{k}}(\tau)d\tau  \nonumber ,
\end{eqnarray}
and
\begin{eqnarray}
G_{\vec{k-q},\vec{q};\vec{k}}^{TN;T}(t) &=& \int_0^t
G^{TT}_{\vec{k-q}}(t-\tau) \label{Gtnt} \\
 && \sum_{a=L,N,H} G^{Na}_{\vec{q}}(t-\tau) 
        \barM^{Ta;T}_{\vec{k-q},\vec{q};\vec{k}} 
G^{TT}_{\vec{k}}(\tau)d\tau  \nonumber .
\end{eqnarray}
In Eqs.~(\ref{Gtlt}) and (\ref{Gtnt}), to leading order in the
wave-vectors,
the vertex $\barM^{TL;T}$ for $a=L$
contributes at Euler order and is imaginary, while the vertices
$\barM^{TN;T}$ and $\barM^{TN;T}$ contribute at dissipative order and
are real.  However, since $G_{\vec{q}}^{LL}(t-\tau)$,
$G_{\vec{q}}^{NN}(t-\tau)$, $G_{\vec{q}}^{HN}(t-\tau)$
and $G_{\vec{q}}^{NH}(t-\tau)$ are real and
$G_{\vec{q}}^{LN}(t-\tau)$,
$G_{\vec{q}}^{NL}(t-\tau)$ and $G_{\vec{q}}^{LH}(t-\tau)$ are purely
imaginary by time-reversal symmetry, the correlation functions
$G_{\vec{k-q},\vec{q};\vec{k}}^{TT;N}(t)$ and
$G_{\vec{k-q},\vec{q};\vec{k}}^{TN;T}(t)$ are
real whereas $G_{\vec{k-q},\vec{q};\vec{k}}^{TL;T}(t)$ is purely
imaginary.  Note that at $t=0$, the expressions for the functions
$G^{21}$ in Eqs.~(\ref{Gttn}), (\ref{Gtnt}) and (\ref{Gtlt})
vanish and the multi-point correlation functions are given exactly.

\subsection{Three-time correlations}

The three-time correlation functions, 
\begin{eqnarray*}
G_{\vec{k-q},\vec{q},\vec{k}}^{TLT}(t_{1},t_2) &=& 
\frac{\saverage{T_{\vec{k-q}}(t_1+t_2)L_{\vec{q}}(t_1)T_{-\vec{k}}}}{\average{N}mk_{B}T} \\
G_{\vec{k-q},\vec{q},\vec{k}}^{TNT}(t_1,t_2) &=& 
\frac{\saverage{T_{\vec{k-q}}(t_1+t_2)\hat{N}_{\vec{q}}(t_1)T_{-\vec{k}}}}{\average{N}mk_{B}T},
\end{eqnarray*}
can be evaluated in a straightforward fashion using the results of the
previous section. In Ref.~\cite{vanZon}, it was shown that the
multi-time vertices $\barM^{lmn}$ reduce to very simple forms to
leading $N$-order, with corrections of order $M/N \sim k_{c}a \approx
10^{-5}$ for systems of moderate density\cite{SLO92}.  Using the reduced forms of
$\barM^{211}$ and $\barM^{112}$ and \eq{generalMT}, the leading
$N$-order expressions for these multi-time functions are,
\begin{eqnarray}
G_{\vec{k-q},\vec{q},\vec{k}}^{TLT}(t_{1},t_2) &=& 
G^{T;TL}_{\vec{k-q};\vec{k},\vec{-q}}(t_{2}) \;
\saverage{L_{\vec{q}}L_{-\vec{q}}} \; G_{\vec{k}}^{TT}(t_{1})
\nonumber \\
&& + G_{\vec{k-q}}^{TT}(t_{2}) \; G^{TL;T}_{\vec{k-q},\vec{q};
\vec{k}}(t_{1}) ,
\label{mtTLT1}
\end{eqnarray}
and
\begin{eqnarray}
G_{\vec{k-q},\vec{q},\vec{k}}^{TNT}(t_{1},t_2) &=& G_{\vec{k-q}}^{TT}(t_{2}) \; \frac{
S(q)}{\average{N}} \; G_{\vec{k}}^{TT}(t_{1}) \label{mtTNT1}\\ 
&& + G^{T;TN}_{\vec{k-q};\vec{k},\vec{-q}}(t_{2}) \;
S(q) \; G_{\vec{k}}^{TT}(t_{1})
\nonumber \\
&& + G_{\vec{k-q}}^{TT}(t_{2}) \; G^{TN;T}_{\vec{k-q},\vec{q};
\vec{k}}(t_{1}) .
\nonumber
\end{eqnarray}
Using the symmetry properties of $G^{lm}(t)$\cite{vanZon},
one can write 
$G^{T;Ta}(t) \saverage{\hat{A}^{a} \hat{A}^{a*}} =\big( G^{Ta;T}(-t)
\big)^{*}$, and
Eqs.~(\ref{mtTLT1}) and (\ref{mtTNT1}) can be expressed in terms of
$G^{21}(t)$ alone as,
\begin{eqnarray}
G_{\vec{k-q},\vec{q},\vec{k}}^{TLT}(t_{1},t_2) &=& 
- G^{TL;T}_{-\vec{k},\vec{q};\vec{q-k}}(t_{2})  \; G_{\vec{k}}^{TT}(t_{1})
\nonumber \\
&& + G_{\vec{k-q}}^{TT}(t_{2}) \; G^{TL;T}_{\vec{k-q},\vec{q};
\vec{k}}(t_{1}) ,
\label{mtTLT2}
\end{eqnarray}
and
\begin{eqnarray}
G_{\vec{k-q},\vec{q},\vec{k}}^{TNT}(t_{1},t_2) &=& G_{\vec{k-q}}^{TT}(t_{2}) \; \frac{
S(q)}{\average{N}} \; G_{\vec{k}}^{TT}(t_{1}) \label{mtTNT2}\\ 
&& + G^{TN;T}_{-\vec{k},\vec{q};\vec{q-k}}(t_{2}) \;
G_{\vec{k}}^{TT}(t_{1})
\nonumber \\
&& + G_{\vec{k-q}}^{TT}(t_{2}) \; G^{TN;T}_{\vec{k-q},\vec{q};
\vec{k}}(t_{1}) ,
\nonumber
\end{eqnarray}
where the time-reversal symmetry properties and the behavior under
complex conjugation of the $G^{21}(t)$
correlation functions has been used.

In Section IV, these expressions will be compared to results from simulations
of a hard sphere system at moderate density.

\section{Simulation method}                              

%
%
%
The dynamics of hard spheres consist of free, rectilinear motion until
the distance between two spheres ($i$ and $j$) becomes equal to their
diameter $a$, at which point an instantaneous collision takes place,
leading to the momentum changes
\begin{eqnarray*}
	{\bf p}_i &\rightarrow& {\bf p}_i 
		- \hat\sigma[({\bf p}_i-{\bf p}_j)\cdot\hat\sigma]
\\
	{\bf p}_j &\rightarrow& {\bf p}_j 
		+ \hat\sigma[({\bf p}_i-{\bf p}_j)\cdot\hat\sigma] ,
\end{eqnarray*}
where the collision normal $\hat\sigma$ equals $({\bf r}_j-{\bf r}_i)/a$
at contact. 

Due to the simplicity of the equations of motion, the dynamical
evolution of the hard sphere system can be computed exactly using an event-driven
procedure in which one calculates the first possible collision of all
spheres under the assumption that no other particles collide.
The phase point of the system is then evolved up to the time of the
earliest of these collisions,  and the process is repeated until the
total desired run time is completed.

Without additional bookkeeping, the number of spheres with which a specific
particle can collide is $N-1$, and hence $O(N)$
calculations of collision times are required for each particle after
it collides. As the
number of collisions per unit time is extensive, the simulation time scale
increases quadratically with the number of particles. 
Considerable improvements in simulation efficiency can be gained using
a division of the system into regions (called cells) and data
structures to optimize the search for the next collision time\cite{R80}. 

To use the cell structure in a simulation, the system of dimension $L_x\times
L_y\times L_z$ is divided into an
integer number of cells of dimension $l_x\times l_y\times l_z$,
where each of the lengths $l_x$, $l_y$ and $l_z$ is no smaller than the diameter of
the hard spheres.  Now, in addition to the collision events between spheres,
the cell in which each sphere is located and
the time at which the particle will leave its cell is recorded. 
This is advantageous because the number of spheres that can collide with a
given sphere before a particle moves out of its cell is proportional to the number of spheres in its
vicinity, i.e., the spheres in the same cell or in one of the $26$
neighboring cells. Using the cell structures, the number of spheres
within the vicinity of
a given particle is of order $O(l_xl_yl_zN/L_xL_yL_z)=O(1)$, provided
the lengths of each cell are on the order of the diameter of the
particles, and hence far fewer collision times of pairs of particles
must be computed after each collision event.  However, the use of
cells comes at the cost of increasing the complexity of the event
driven simulation since after a crossing event for a given particle, the
collision times of the given sphere with spheres that previously were
not in its vicinity must be considered, and, if necessary, the first stored
collision event adjusted. In addition, the next crossing time in the same
direction is re-calculated.  Similarly,
after a collision event between two particles, new collisions within
the same cell as well as the new cell-crossing times must be
calculated for the particles involved in the event.

Even though the calculations after a crossing or a collision event are of
$O(1)$ when many cells are used, it is still necessary to
search the event list of each sphere to find the earliest event in the
simulation.
If the spheres are simply stored in a linear array, this implies a
look-up time that scales linearly with $N$, and the algorithm scales
as $N^2$ as before, though with a considerably lower prefactor than without the
cell structure.
If, on the other hand, the spheres are stored in a binary tree,
ordered according to their first event, the search for the first event
scales as the
logarithm of the number of elements, which in our implementation is
$N$. Deleting an element from the tree is an $O(1)$ operation, while the insertion of new
elements into the tree requires a tree search, which scales as $\ln N$.  
Since the number of crossing and collision events is extensive, the
algorithm scales as $N\ln N$, and the overall speed up of the algorithm over a simple event-based
simulation behaves as $N/\ln N$. It should be noted,
however, that the cell structure reduces the number of collisions to be
considered to a large extent, so the prefactor is also quite reduced.

There is some flexibility in selecting the size of the cells to be used in
the simulation.  Larger cells require fewer crossing times to be
calculated at the expense of increasing the number of collisions which
must be computed within each cell. As Rapaport has noted\cite{R80},
the optimal choice of the dimensions of the
cell for systems of low density is intermediate between the size of the full system and the
diameter of the hard spheres, whereas the
smallest possible cells make for the fastest simulation for higher
densities.  In the simulations reported in the next section, the
optimal length of cells was found to correspond roughly with the diameter
of the hard sphere particles.

\section{Results and discussion}
In this section, the mode coupling expressions for the higher-order correlations
functions given in section II are compared to those obtained from
event-based molecular dynamics simulations in the micro-canonical
ensemble at an inverse temperature $\beta=3$.  
The size of the periodic system in the simulation was chosen to be
$L_x=L_y=L_z=15.7526$, such that for $N=1382$ hard sphere particles of
diameter $a=1$, the reduced density $\rho^{*}=0.25$ ($\rho/\rho_c$
where $\rho_c$ is the density at close-packing) and the
magnitude of the smallest wave-vector $k_0a=2\pi a/L_x=0.398867$
coincide with one of the cases in Ref.~\cite{KCdS87}.  
A total number of $15^{3}=3375$ cells were used, leading to a
collision rate (including data collection) of roughly $3.2 \times 10^{6}$/hour of CPU time on a 
$600$ MHz digital 21164 processor.  The event-dynamics simulations were run on a
$9$ nodes of a $30$-node ``Beowulf'' cluster for a total of $4,100$ CPU hours, where
each node carried out $3,750$ short molecular dynamics trajectories of
approximately $403,500$ collisions.  The initial configuration of the system
for each of the individual runs was randomly chosen using a simple
rejection method.  In all results reported below, time is
expressed in dimensionless units $t/t_{c}$, where $t_c$ is the mean
collision time calculated from the simulation.  At the density and
temperature of the simulation, the
mean collision time is roughly half the time $t_{m}$ it takes a
particle to move over a distance equal to its diameter $a$ ($t_c/t_m \approx 0.412$).

To evaluate time correlation functions in the simulation, 
the values of the linear densities $A_{\vec{k}}(t)$ were calculated for
a set of wave-vectors at $M+1$ fixed time intervals $t=0,\Delta
t,2\Delta t,\ldots$ 
and stored in an array $A[k][i]$, where the index $k$ runs over the
wave vector indices and $i$ runs from $0$ to $M$.  In all molecular
dynamics runs, the time interval $\Delta t/t_m=0.15$, and $M=400$.
Two-point, two-time correlation functions for a given time interval
were accumulated on the fly by storing the product of accumulated
arrays $A\big[k\big]\big[(t/\Delta t)\mbox{mod }M\big] *
A^*\big[k\big]\big[\{(t-s)/\Delta t\}\mbox{mod }M\big]$ in an array
for the correlation function $\saverage{A_k(s)A_{k}^*}$ for all relevant values of $s$. At the end
of the run the result was divided by the number of points accumulated.
Multi-point and multi-time correlation function are
evaluated in an analogous fashion.

Good statistics are difficult to obtain for the higher-order correlation
functions since the functions are the average of a product of multiple
factors of the linear densities $A_{\vec{k}}$.  For example, the
three-point correlation functions are constructed by
averages of quantities which are typically on the order of $N^3$, whereas the
final average itself is of $O(N)$. In order to optimize the sampling, 
many relatively short runs of duration $R=4M\Delta t$ were performed and averaged on the
fly. The strategy of using many short runs seems to be better than the
alternative of performing a single long
run of equal total length, perhaps because it reduces the effect of
abnormally large points which contaminate the signal for a long time.

Further improvement of the statistics of the calculated correlation
functions is possible by exploiting the
isotropy of the system. To simplify the comparison between theoretical predictions and the
simulation results, all wave-vectors
were taken to be co-linear along the $\hat{x}$ axis so that $\vec{k}
\cdot \vec{q}= kq$, where $k=|\vec{k}|$ and $q=|\vec{q}|$.  Since the
wave-vectors $\vec{k}$ and $\vec{q}$ are parallel,
the quantities
$\saverage{A_{k\hat{x}-q\hat{x}}(t) A_{q\hat{x}}(t)A^*_{k\hat{x}}}$,
$\saverage{A_{k\hat y-q\hat y}(t)A_{q\hat y}(t)A^*_{k\hat y}}$ and
$\saverage{A_{k\hat z-q\hat z}(t)A_{q\hat z}(t)A^*_{k\hat z}}$ can be
computed from the simulation in a periodic, cubic simulation box and averaged to obtain improved
statistics.  In addition, 
for many of the correlation functions considered here, such as 
$\saverage{T_{k-q}(t) N_{q}(t)T^*_{k}}$,
the number of points used to calculate the higher-order correlation
functions can be effectively doubled by averaging over the transverse
directions $\hat{y}$ and $\hat{z}$.

The estimation of statistical uncertainty in the simulation data is
problematic as it involves constructing an auto-correlation function
for each point measured in the time-correlation
function\cite{timeseries}. 
Such a procedure is both memory and computationally intensive, and
slows down the simulation dramatically.  In fact, most of the
computational time of the simulation is spent accumulating data and
calculating the correlation functions rather than performing the
molecular dynamics.
We therefore adopt a simpler approach to estimate the error using the
symmetry properties of the correlation functions. From reflection symmetry, it
follows that all correlation functions are either real or imaginary\cite{vanZon}. For a
real correlation function, the imaginary part vanishes and hence the
imaginary part calculated from the simulation gives a rough estimate of the error in the real 
part, as both are calculated from the same configurations and involve
terms of similar structure. To approximate the statistical
uncertainty for a real correlation function, a histogram of the values of the imaginary part is
constructed to determine the interval of values containing $96\%$ of the points. The size of
this interval provides an estimate of the error in the real part,
taken to be constant for all times of the correlation function. Such
an approximation seems reasonable given that
the variations in the imaginary part in the simulations are observed
to be relatively constant over the total time interval $M\Delta t$.
For an imaginary correlation function, the analogous procedure is done
using the variations in the real part.

The simulation results for the two-point, two-time correlation
functions were checked against generalized Enskog theory
results\cite{KCdS87}. The statistical uncertainty in the normalized
correlation functions (as obtained by the procedure above) are quite
small (of the order of $0.001$).
The numerical value for the shear viscosity extracted from the exponential
decay of the auto-correlation function of the transverse velocity
$T_{\vec{k}}$ was compared to the kinetic theory prediction for this
quantity\cite{Chapman} and excellent agreement was observed.

The time convolution integrals in the mode coupling expressions for
the higher-order correlation functions were evaluated by numerically
integrating data for the two-point, two-time correlation functions
$G_{k}^{ab}(t)$ obtained directly from the simulation.  Since the
error bars of the $G_{k}^{ab}(t)$ are very small, the level of
uncertainty in the theoretical prediction for the higher order
corelation functions is negligible in comparison to the uncertainty in
the simulation data for the higher-order correlation function.
Furthermore, no significant differences were noted in the convolution integrals
calculated using the simulation data and calculated from high quality
functional fits of the integrands.  In principle, one could also use
the hydrodynamic forms for all two-point, two-time correlation
functions in combination with an accurate equation of state and
kinetic theory results for the transport coefficients, but since the
simple correlation functions were obtained with great accuracy in the
simulation, the actual data was used.

As described in the appendix, the dissipative part of the vertices for
$\barM^{TN;T}$, $ \barM^{TH;T}$ and $\barM^{TT;H}$ have free
parameters $v_{n}$, $v_{h}$ and $v_{th}$ which must be fitted to the
data if the dissipative contributions are to be included in the
predictions for both the multi-point and multi-time correlation
functions.  In practice, this is accomplished by selecting 
particular wave-vector magnitudes $k$ and $q$ and fitting the parameters
according to the simulation results.  This procedure is illustrated in
Fig.~1 for the multi-point correlation function 
\[
C^{TL;T}(t) = \saverage{T_{k-q}(t) L_{q}(t)
T_{-k}}/\saverage{N}mk_{B}T
\]
for wave-vectors $k=k_0$ and $q=2k_0$.  
Note that
although the dissipative contribution to the overall correlation
function in Eq. (\ref{Gtlt}) depends on the two parameters $v_n$ (N-coupling) and $v_h$
(H-coupling), these parameters can be uniquely determined since the asymptotic time
behavior is determined entirely by the N-coupling contribution.  Once
this parameter is set (here, $v_n=-0.18$), $v_h$ can be determined by
fitting the height of the first peak (found to be $v_h=0.90$).  
A similar procedure is used for the $\barM^{TT;H}$ vertex
in Eq. (\ref{Gttn}) which is relevant for the correlation function 
\[
C^{TT;N}(t)=\saverage{T_{k-q}(t)
T_{q}(t) \hat{N}_{-k}}/S(k), 
\]
and it is found that $v_{th}=-0.62$.
Note that it is, in fact, the additional couplings which arise at
dissipative order that account for the slow decay of the three-point
correlation function in Fig.~1.  It is therefore quite apparent that
ordering of terms using the wave-vector must be done carefully for the
system under consideration, since contributions which appear at
higher-orders of wave-vector can actually dominate lower order terms.

With the coupling parameters fixed by the fitting procedure, one can
then compare the simulation results with the theoretical predictions
for arbitrary wave-vector combinations.  In Fig.~2, the simulation and
theoretical predictions for the multi-point
correlation functions $C^{TL;T}(t)$ and $C^{TT;N}(t)$
are shown as a function of time for a
number of wave-vector combinations.  The remarkable agreement between
the simulation results and the theoretical predictions of both
three-point correlation functions over all time-regimes and
wave-vector combinations is a clear indication that the formulation of
the mode-coupling theory is sound.

It is interesting to see how the theoretical predictions of the
present formalism compare to those obtained from a mode coupling
theory in which Gaussian statistical behavior is assumed in the
multi-linear basis set.  This type of assumption roughly corresponds
to Kawasaki's original formulation of mode coupling theory\cite{K70}, which is
based upon a non-linear Langevin equation with Gaussian noise (or
fluctuating forces).  Such a Gaussian theory for the multi-point
correlation functions differs from the present
formulation in two significant ways: First, since the subtractions in
the multi-linear basis set (see \eq{Qset}) involve static three-point correlation
functions of the linear densities which vanish under the assumption of
Gaussian statistics, the Gaussian theory neglects terms of the form 
\begin{eqnarray*}
 \saverage{\hat{A}_{\vec{k}}(t)
\hat{A}_{-\vec{k}}}
\cdot K_{11}^{-1} \cdot \saverage{ \hat{A}_{-\vec{k}}
\hat{A}_{\vec{k-q}}\hat{A}_{\vec{q}}} 
\end{eqnarray*}
which appear, for instance, in the mode coupling expression 
for $C^{TT;N}(t)$.  These terms make an
important contribution to the multi-point correlations functions on
all time scales, and particularly for short times.  Second,  since the
subtraction terms vanish in the Gaussian theory, the coupling vertices
are significantly affected.  For example, looking at the Euler order
contributions to the vertex $\barM^{TT;N}$, the vertex in the Gaussian
approximation becomes $V^{TT;N}$, according to 
\begin{eqnarray}
\barM^{TT;N} &=& \average{ \dot{Q}^{TT}_{k-q,q} \hat{N}_{-k}} / S(k)
\nonumber \\
&=& \average{\dot{ \big( T_{k-q}T_{q} \big) } \hat{N}_{-k}} / S(k)
\equiv V^{TT;N}.
\end{eqnarray}
Similar differences between the vertices $V$ in the Gaussian
approximation and the $\barM$ vertices appear in the dissipative parts of the
$\barM^{TN;T}$ and $\barM^{TH;T}$ (see Tables I and II). 

In order to assess how each of these differences affects three-point
correlation functions, we once again consider the correlation functions 
$C^{TT;N}(t)$ and $C^{TL;T}(t)$.  The first correlation function differs not only
in the explicit form of the coupling vertices but also in the form of
the expressions due to the subtraction terms in the basis set. 
The second correlation function, on the other hand, vanishes at $t=0$
by symmetry and does not contain
contributions which are directly proportional to two-point, two-time correlation functions (see
Eq. \ref{TLTfinal}).  For this correlation function, the differences between the Gaussian and full mode
coupling theory arise solely due to differences between the Gaussian
vertices $V^{TN;T}$, $V^{TH;T}$ and their full counterparts.
In Fig.~3, the Gaussian and full mode coupling expressions are compared to the
simulation data for the wave-vectors $k=k_0$, $q=3k_0$.  
From these plots, it is clear that the Gaussian theory poorly predicts
the time-dependence of $C^{TT;N}(t)$ on all time scales, and also gives worse results for
the correlation function $C^{TL;T}(t)$.
Similar behavior can be seen for other wave-vector combinations.

Turning now to the multi-time correlation functions, the
simulation results and theoretical predictions for the 
multi-time correlation functions $G^{TLT}(t_{1},t_2)$ and $G^{TNT}(t_{1},t_2)$ for several
different wave-vectors are plotted in Fig.~4 as a function of time $t$ for the time combinations $(t_1,t_2)$ of
$(t,t)$, $(t,3t)$ and $(3t,t)$.  The excellent agreement between the
full mode coupling theory and simulation results strongly suggests that the assumptions discussed
at length in Ref. \cite{vanZon} of what
determines whether a correlation function decays quickly or not are
appropriate.  These assumptions 
are necessary to obtain mode coupling equations which are local in
time.  

Note that the time symmetry properties are evident in the two graphs
in the first row of Fig.~4, which correspond to the wave-vectors
$k=k_0$, $q=2k_0$.  For these wave-vectors, the time symmetries can be
obtained by noting that 
\begin{eqnarray*}
\saverage{T_{k-q}(t_{1}+t_2) A^{a}_{q}(t_1)T_{-k}} &=&
\saverage{T_{k-q}(t_2) A^{a}_{q}T_{-k}(-t_1)}
\end{eqnarray*}
since the equilibrium distribution function is stationary.  Inverting
all time arguments and using the properties of the densities under
time-reversal, one obtains
\begin{eqnarray*}
\saverage{T_{k-q}(t_2) A^{a}_{q}T_{-k}(-t_1)} &=&
\gamma_{a} \saverage{T_{k-q}(-t_2) A^{a}_{q}T_{-k}(t_1)},
\end{eqnarray*}
where $\gamma_a=1$ for $a=N,H$ and $\gamma_a=-1$ for $T,L$.  When
$k-q=-k$, we find that
\[
\saverage{T_{-k}(t_1+t_2)L_{2k}(t_1)T_{-k}}=-
\saverage{T_{-k}(t_1+t_2)L_{2k}(t_2)T_{-k}}, 
\]
which implies this
correlation function is anti-symmetric under interchange of $t_{1}$
with $t_{2}$ (and therefore vanishes when $t_{1}=t_{2}$), while 
$\saverage{T_{-k}(t_1+t_2)L_{2k}(t_1)T_{-k}}$ is symmetric under the
exchange of $t_{1}$ and $t_{2}$.  It is reassuring, though not
surprising, that the mode coupling theory respects these time-reversal properties.

One may also calculate the multi-time correlation functions
via Eqs.~(\ref{mtTLT2}) and (\ref{mtTNT2}) using the simulation data
for the multi-point functions $G^{TN;T}$ and $G^{TL;T}$.
However, since the mode coupling results for these functions are
already in excellent agreement with the simulation data, the
improvement obtained using the simulation results for the $G^{21}$ is
generally statistically negligible.  Furthermore, the simulations to
calculate the multi-point functions are computationally-intensive
compared to calculations of the two-point functions.  It is therefore
far easier to generate predictions with small statistical
uncertainties using the mode coupling theory expressions for the
multi-point functions.

Since the mode coupling formalism relates the multi-time correlation
functions to multi-point correlation functions, 
the deficiencies in the Gaussian theory
for the three-point functions are carried over to the predictions for three-time functions.
This point is confirmed by the difference in the behavior of the Gaussian versus full
mode coupling theory results for the multi-time correlation
functions shown in Fig.~5.  Once again, the Gaussian theory
predictions for multi-time correlations is qualitatively incorrect on
all time scales, and particularly so for correlation functions which
do not vanish when $t_1=t_2=0$.  Furthermore, as might be expected
from the discussion above of the dissipative contribution to the three-point
correlation functions, the inclusion of the additional couplings
arising at dissipative order is essential if quantitatively accurate
predictions for the multi-time correlation functions is desired.

In principle, in the limit of very small wave-vectors, one might
expect that the additional couplings in the higher-order correlation arising from the dissipative
part of the vertices become less important and may be neglected. 
In fact, this is not always the case since the overall order in
wave-vector of the various terms in the expressions for $G^{21}$ is determined
by a wave-vector dependent pre-factor (the vertex) multiplied by 
the time-convolution of two-point, two-time correlation functions.
The time-convolution of functions such as $G_{k}^{LN}(t)$, which
vanish as $k \rightarrow 0$, can give additional factors of the
wave-vector.  Thus, for instance, the contribution from
the first term in Eq.~(\ref{TNTfinal}) (with a vertex of Euler order)
is in fact of the same order of magnitude as the contribution from
the last two terms, which involve vertices of dissipative order in the
hydrodynamic limit.

To obtain smaller wave-vectors in a simulation to numerically check these
considerations for dense systems in which the mode coupling effects
are important,  one would need to simulate larger systems with more
particles.  There are two difficulties with the simulation method
applied to larger systems which make it difficult to obtain good
statistics for the higher-order correlation functions. First,
since the use of cells is memory intensive and the optimal number of
cells scales as the cube of the length of the system, one must utilize a cell-structure for
the simulations which is not optimal, leading to a reduction in
simulation efficiency.  Second, the quality of the statistics for the higher-order
correlation functions decreases essentially as the square of the
number of particles.  It is therefore computationally challenging
to obtain accurate simulation results for the higher-order correlation
functions for larger systems.

\section{Summary and conclusions}

In this paper, the predictions for higher-order correlation functions
based on the mode coupling formalism developed in
Ref.~\cite{vanZon} were evaluated in the hydrodynamic limit for a hard
sphere system at moderate densities and compared to simulation
results.  It was demonstrated that the mode coupling theory yields
excellent results for all higher-order correlation functions provided
that dissipative as well Euler order vertex coupling terms are included in
the theory.  The good agreement between the theoretical predictions
and the simulation results confirms that the assumptions underlying the mode coupling theory of how slow
and fast decay of arbitrary densities can be separated in a systematic
fashion are quite reasonable.  

In contrast to some mode coupling theories of simple
liquids\cite{K70,R81,Gotze75,Bosse78},  the present mode coupling
theory includes all multi-linear densities in the set of slow
variables, does not neglect corrections to the ``factorization''
approximation, and does not assume Gaussian statistical properties of
the random noise or fluctuating force.  As the formalism allows exact
expressions to be obtained for all correlation functions in the
thermodynamic limit, it provides a systematic way to examine the
various assumptions which must be made in order to predict the
time-dependence of simple and higher-order correlation functions, or to
form comparisons with other theories.  Along these lines, it was
demonstrated that the ``non-Gaussian'' behavior of the random noise is
important for the proper description of the multi-point correlation
functions on all time scales.  In particular, the Gaussian theory
for three-point functions leads to over-simplified coupling vertices which have significant
quantitative consequences and, more importantly, neglects important
couplings to linear densities.  Since the mode coupling theory
expresses the multi-time correlation functions in terms of two-time,
higher-order correlation functions, the Gaussian theory has similar
deficiencies in describing the three-time correlation function of
linear densities.

The calculation of higher-order correlation functions of extensive
linear densities in the hydrodynamic regime at low to intermediate
densities is computationally intensive.  The poor statistics obtained
from the simulation arises from averaging quantities of order $N^{3}$
to obtain a signal of order $N$.  However, since densities of tagged particles
do not scale with the number of particles, higher-order correlation
functions of tagged particle densities should not suffer from this problem.
The extension of the mode coupling
theory of higher-order correlation functions to non-extensive
densities of tagged particles is straightforward, and will be
presented in a future publication.  

It is obviously desirable to apply the mode coupling formalism to
dense and super-cooled liquids where correlation functions exhibit more
complicated time behavior.  In dense systems, there is compelling
evidence\cite{KCdS87} which suggests that the eigenmode spectrum of the 
Liouville operator for simple liquids changes, and a generalized
``heat'' mode becomes long-lived even at fairly large wave-vectors.
At large wave-vectors, this mode roughly corresponds to a
self-diffusion mode\cite{dSC80} which is slow in dense liquids due
to particle caging effects.  Within the mode coupling formalism, the emergence of
this short-wavelength collective mode implies that the cut-off
wave-vector $k_{c}$ for the heat mode becomes on the order of inverse molecular length
scales.  Under
these circumstances, the mode coupling correction terms to the
expressions for the higher-order correlation functions are not
expected to be small and must be considered.  Appropriately-defined higher-order correlation
functions may be quite useful in examining the microscopic origins of
complex relaxation behavior and dynamical heterogeneities.  To this
end, one may examine
the higher-order correlation functions at much larger wave-vectors
using a mode coupling theory in which the modes
forming the basis set for the long-time behavior are associated with 
physical processes on these length scales.  In fact,
the structure of the mode coupling theory suggests that measures of
dynamical heterogeneity based on multi-point
correlation functions\cite{Glotzer} are quite closely related to
measures based on multi-time correlation functions\cite{Heueretal}.  These
issues are currently being pursued.

\section*{Acknowledgments}
This work was supported by a grant from the
Natural Sciences and Engineering Research Council of Canada and funds
from the Premier's Research Excellence Award. 

\appendix
\section{Evaluation of the vertices}
\label{appendix} 
\newcommand{\pbracket}[2]{\{#1\,,#2\}} 

In this appendix, all vertices used to formulate numerical
predictions for higher-order correlation functions are given for the
sake of completeness.  To leading order in the wave-vectors, all
vertices are either of Euler order (order $k$), or of dissipative order
(order $k^2$).  Since the second term in the expression for the
vertices in \eq{vertex} involves two time derivatives of hydrodynamic
densities, it is at least of quadratic order in the wave-vectors.
Therefore the Euler order of any vertex is given by the static
correlation function (first term) of \eq{vertex}.  This static
correlation function is imaginary and an odd function of wave-vector.
The leading order of a vertex of quadratic order in wave-vector is
therefore given by the second term in \eq{vertex}.  For the hard
sphere system, all static correlation functions in the zero wave-vector
limit can be evaluated exactly if
the radial distribution function at contact $g(a)=\chi$ is known. The calculation of
the vertices at Euler order is facilitated by considering the identity, valid
in the canonical and grand canonical ensembles,
\begin{equation}
	\saverage{\dot A \,B} = \beta^{-1}\saverage{\pbracket{A}{B}},
\eql{poisson}
\end{equation}
which links the time derivative to a Poisson bracket of the densities.
It follows from $\dot{A}=\{A\,,{\cal H}\}$ and from the form of the
distribution function:
\[
	\int \pbracket{A}{{\cal H}}B e^{-\beta {\cal H}}d\Gamma
	= \int \left[ 
	\dd{A}{q}B\dd{{\cal H}}{p} - \dd{A}{p}B\dd{{\cal H}}{q}
	\right] e^{-\beta {\cal H}} d\Gamma,
\]
which, by partial integration, yields
\begin{eqnarray*}
&&	\beta^{-1}\int \left[
		\dd{\ }{p}\left(B\dd{A}{q}\right)
		-\dd{\ }{q}\left(\dd{A}{p}B\right)
	\right] e^{-\beta{\cal H}} d\Gamma
\\&&
	= \beta^{-1}\int \pbracket{A}{B} e^{-\beta{\cal H}} d\Gamma. 
\end{eqnarray*}

To evaluate the higher-order correlation functions in the text, the vertices $\bar M^{TT;L}$ and
$\bar M^{TL;T}$ are needed. The latter is the simplest, as 
$Q_2^{TL}=T_{\vec k-\vec q}L_{\vec q}$, so
\begin{eqnarray*}
\bar M^{TL;T}_{\vec k-\vec q,\vec q;\vec k}
&=& \beta^{-1}\saverage{\pbracket{T_{\vec k-\vec q}L_{\vec q}}{T_{\vec k}^*}}/
 \saverage{T_{\vec k}T^*_{\vec k}}
\\
&=& ik\beta^{-1}\saverage{T_{\vec k-\vec q}T^*_{\vec k-\vec q}}/
 \saverage{T_{\vec k}T^*_{\vec k}} = ik\beta^{-1},
\end{eqnarray*}
where we used 
$\pbracket{AB}{C}=A\pbracket{B}{C}+\pbracket{A}{C}B$, 
and the facts that $\pbracket{T_{\vec q}}{T_{\vec k}^*}=0$ and 
$\pbracket{L_{\vec q}}{T_{\vec k}^*}=ikT^*_{\vec k-\vec
q}$.

It is straightforward to show that $Q_2^{TT}=T_{\vec k-\vec q}T_{\vec
q}-\frac23mE_{\vec k}$, so using the above result for $\{L,T\}$, we obtain
\begin{eqnarray*}
\bar M^{TT;L}_{\vec k-\vec q,\vec q;\vec k} &=& 
	-ik\beta^{-1} 
	- \sfrac23 \saverage{\pbracket{E_{\vec{k}}}{L_{\vec{k}}^*}} .
\end{eqnarray*}
The energy can be split into a kinetic and a potential part. The
kinetic contribution to $M^{TT;L}$ is easily calculated, and
$M^{TT;L}$ can be expressed as,
\begin{eqnarray*}
\bar M^{TT;L}_{\vec k-\vec q,\vec q;\vec k} &=& 
	i\sfrac23 k\beta^{-1} 
	- \sfrac23 \saverage{\{E^{\text{pot}}_{\vec{k}},L_{\vec{k}}^*\}}.
\end{eqnarray*}
The second term on the right-hand side of the equation above can be
evaluated by noting that
\begin{eqnarray*}
	\{E^{\text{pot}}_{\vec k},L^*_{\vec k}\} 
	&=& \sfrac12
		\sum_{j\neq m}\left[e^{i\vec k\cdot\vec r_{mj}}-1\right]
		\hat x\cdot\dd{}{\vec r_j}V(r_{mj})
\\
	&=& -\sfrac12 \sum_{j\neq m} i(\vec k\cdot\vec r_{mj} )
		\hat x\cdot\dd{}{\vec r_m}V(r_{mj})
		+
		O(k^2) ,
\end{eqnarray*}
where $\vec r_{mj}=\vec r_m-\vec r_j$, which implies
\begin{eqnarray*}
	\saverage{\{E^{\text{pot}}_{\vec{k}},L_{\vec{k}}^*\}}
	= -\sfrac12 ik \average{N} \rho\int r (\hat x\cdot\hat r)
	\hat x\cdot(\dd{}{\vec r}V(r)) g(r) \,d\vec r,
\end{eqnarray*}
where $g(r)$ is the radial distribution function.
Performing the angular integration and writing $g(r)=h(r)e^{-\beta
V(r)}$ so that a partial integration can be performed, leads to
\[
	\saverage{\{E^{\text{pot}}_{\vec{k}},L_{\vec{k}}^*\}}
	= -ik\frac{2\pi\rho \average{N}}{3\beta}\int_a^\infty 
	\!\!\dd{}{r}(r^3 h(r)) dr
	= -ikb\rho N\chi,
\]
where $b\equiv 2 \pi a^{3}/3$, and $\chi$ is the radial distribution
function at contact.  $\chi$ can be estimated using the Carnahan-Starling equation of
state\cite{Carnahan} and the expression for the pressure
$p$
of a hard-sphere system,
\begin{eqnarray*}
\frac{\beta p}{\rho}
&=&
\frac{1+\bar{\eta}+\bar{\eta}^{2}-\bar{\eta}^{3}}{(1-\bar{\eta})^{3}}
\\
&=& 1 + b\rho \chi,
\end{eqnarray*}
where $\bar{\eta}$ is the packing fraction given by $\bar{\eta}=\pi
\rho a^{3}/6$.

Combining all terms, one obtains
\[
	M^{TT;L}_{\vec{k}-\vec{q}\vec{q};\vec{k}}
	= i\sfrac23 kp/\rho .
\]

Turning now to the calculation of the dissipative part of vertices, 
their specific wave-vector dependence is
determined as follows:
The derivative of a conserved density can be written as 
\[
	\dot{\hat{A}}^a_{\vec k} = i\vec k\cdot J^a_{\vec k}
\]
where $J^a_{\vec k}$ is the current associated with the hydrodynamic
variable $a$. What is needed in \eq{flucforce} is the {\em dissipative
current} $j^a_{\vec k}\equiv (1 -{\cal P}) J^a_{\vec k}$. Looking
first at the vertex $\bar M^{TH;T}$, using
\[
	Q^{TH}_{\vec k-\vec q,\vec q}
	= T_{\vec k-\vec q}H_{\vec q} - 
	\frac{\saverage{T_{\vec k-\vec q}H_{\vec q}T_{-\vec
	k}}}{mNk_BT}
	T_{\vec k} 
	= T_{\vec k-\vec q}H_{\vec q} +\sqrt{\sfrac23} T_{\vec k}
,\]
the vertex can be expressed as
\begin{eqnarray}
	\bar M^{TH;T}_{\vec k-\vec q\vec q}
&=&
-k_x(k_x-q_x) v_h
-k_x q_x v_{h}'
\nonumber\\&&
-k^2_x
\sqrt{\sfrac23}\int_0^\infty
\frac{\baverage{j^{T}_{\vec k}(t) j^T_{-\vec k}}}
{m\saverage{N}k_BT}
dt ,
\eql{MTHT}
\end{eqnarray}
where 
\begin{eqnarray}
v_h &=& \int_0^\infty
\frac{\baverage{J^T_{\vec k-\vec q}(t)H_{\vec q}(t)j^T_{-\vec{k}}}}
{m\saverage{N}k_BT}dt \nonumber \\
v_{h}' &=& \int_0^\infty
\frac{\baverage{T_{\vec k-\vec q}(t)J^{H}_{\vec q}(t)j^T_{-\vec k}}}
{m\saverage{N}k_BT} dt.
\eql{parameters}
\end{eqnarray}
Similar expressions can
be obtained for the parameters $v_{n}$ and $v_{th}$ appearing in the 
$M^{TN;T}$ and $M^{TT;H}$ vertices (see Table II). 
To obtain the leading behavior for small wave-vectors, the
wave-vectors in the integrals can be set to zero, and
the projected dynamics Liouvillian in the exponent in
\eq{flucforce} can be replaced by the full Liouvillian.
Then, the Green-Kubo expression for the 
viscosity $\eta$ can be recognized in the last term of \eq{MTHT},
\begin{equation}
\eta = \frac{\beta}{V} \int_0^\infty\baverage{j^{T}(t)
	j^T}dt.
\eql{eta}
\end{equation}
For the viscosity $\eta$ and the heat conduction $\lambda$ (which
figures in the expression for $\bar M^{TT;H}$), we take the Enskog
expressions\cite{Chapman},
\begin{eqnarray*}
	\eta &=& \eta_{0} b\rho \left( \frac{1}{b\rho \chi}+\frac{4}{5}
+0.7614 \; b\rho \chi \right)
,\\
	\lambda &=& \lambda_{0} b \rho \left( \frac{6}{5}+\frac{1}{b
\rho \chi}+0.7574 \; b \rho \chi     \right),
\end{eqnarray*}
where the Boltzmann value of the shear viscosity $\eta_{0}$ and thermal
diffusivity $\lambda_{0}$ are given by 
\begin{eqnarray*}
\eta_0 &=& \frac{5}{16 a^{2}} \sqrt{ \frac{m}{\beta \pi}} \\
\lambda_0 &=& \frac{75}{64a^{2}} k_{B}\sqrt{ \frac{m}{\beta \pi}}.
\end{eqnarray*}
For the particular parameters of the simulation, it was checked by
studying the decay of simple correlation functions that the Enskog
expression are accurate.

In principle, integrals of time-correlations functions of
products of two currents and a density, as in the expression 
for $v_{h}$ and $v_{h}'$ in \eq{parameters}
can be written in the hydrodynamic limit in terms of transport coefficients and
derivatives
of transport coefficients with respect to thermodynamic quantities
like the temperature and chemical potential.  Dissipative
contributions such as these have already been evaluated by
Lim\cite{LimThesis} in the zero
wave-vector limit in the context of generalized hydrodynamics.
In fact, the expression for $v_{h}'$ can be related to the
viscosity\cite{LimThesis} as
\[
v_{h}' = \sqrt{ \frac{2}{3} } \frac{ \eta }{m \rho},
\]
whereas $v_h$ can be expressed in terms of the viscosity and the
derivatives of viscosity with respect to the temperature and chemical
potential.  Using the form for $v_{h}'$, $M^{TH;T}_{\vec k-\vec q,\vec
q;\vec k}$ 
can be written as
\begin{equation}
\bar M^{TH;T}_{\vec k-\vec q,\vec q;\vec k}
= - k_x(k-q_x) \left[ v_h - \sqrt{\frac23}\frac{\eta}{m\rho}\right].
\end{equation}
Since it is not known
how well the derivatives with respect to temperature and energy of the
approximate kinetic theory expressions for the transport coefficients
correspond to their actual
values, $v_{h}$, $v_{n}$ and $v_{th}$ are taken as free parameters
which will be fitted from simulation data.

The expressions for the vertices that are needed in the
text are listed in Tables I and II. Also tabulated are the
vertices one would obtain from a Gaussian theory in which static
three point correlation functions are set to zero. These Gaussian
vertices are denoted by $V_{21}$.

\ecols
\newpage
\begin{center}
Figure Captions
\end{center}
\noindent
Figure~1.  The fitting procedure for the three-point correlation function 
$C^{TL;T}(t)$ for the
wave-vectors $k=k_0$, $q=2k_0$, where $k_0 a=0.398867$.  The
un-connected circles are the simulation results, the solid line is the
full mode coupling results, the dotted
line is the mode coupling results with Euler vertices, and the
long-dashed and dot-dashed lines represent the contributions from the
$N$-dissipative and $H$-dissipative vertices, respectively.  For
clarity, the statistical uncertainties in all quantities have been omitted.

\vspace{.1in}
\noindent
Figure~2. The correlation functions $C^{TT;N}(t)$ (left panels)
and $C^{TL;T}(t)$ (right panels) as a function of reduced time at
various wave-vectors.  In the top row, the wave-vector arguments are
$k=k_0$, $q=2k_0$ (open un-connected circles: simulation results, solid line: MCT
prediction), and $k=2k_0$, $q=k_0$ (open un-connected squares: simulation results, dotted line: MCT
prediction).  In the middle row, the wave-vector arguments are
$k=k_0$, $q=3k_0$ (open un-connected circles: simulation results, solid line: MCT
prediction), and $k=3k_0$, $q=k_0$ (open un-connected squares: simulation results, dotted line: MCT
prediction).  In the bottom row, the wave-vector arguments are
$k=2k_0$, $q=3k_0$ (open un-connected circles: simulation results, solid line: MCT
prediction), and $k=3k_0$, $q=2k_0$ (open un-connected squares: simulation results, dotted line: MCT
prediction).  For
clarity, the statistical uncertainties in all quantities have been omitted.

\vspace{.1in}
\noindent
Figure~3.  The full mode coupling theory (MCT), Gaussian MCT, Euler
order MCT predictions and simulation data for the correlation
functions $C^{TT;N}(t)$ (left panel) and $C^{TL;T}(t)$
(right panel) at wave-vectors $k=k_0$ and $q=3k_0$.  In both panels, the un-connected circles are the
simulation data, the solid, dotted and dashed lines represent the full
MCT theory, the Euler order MCT theory, and the Gaussian MCT theory
results, respectively.  The error estimates represent $96\%$
confidence intervals.  Note that the
Gaussian MCT theory is qualitatively incorrect on all time scales for
$C^{TT;N}(t)$.  

\vspace{.1in}
\noindent
Figure~4.  The multi-time correlation functions
$G^{TNT}(t_{1},t_{2})$ and $G^{TLT}(t_1,t_2)$ as a
function of reduced time for various wave-vector combinations.  In all
panels, the un-connected dots, $\times$'s and triangles correspond to the simulation
results for the time arguments $t_1=t$, $t_{2}=t$, $t_{1}=3t$,
$t_{2}=t$ and $t_1=t$, $t_2=3t$, respectively.  The solid, dashed, and
dotted lines correspond to the respective mode coupling predictions.  The results in the
top, middle and bottom rows are for the wave-vector arguments $k=k_0$,
$q=2k_0$, $k=k_0$, $q=3k_0$, and $k=2k_0$, $q=k_0$.  For
clarity, the statistical uncertainties in all quantities have been omitted.

\vspace{.1in}
\noindent
Figure~5.  The full mode coupling theory (MCT), Euler
order MCT, and Gaussian MCT predictions and simulation data for the multi-time correlation
functions $G^{TNT}(t_1,t_2)$ (left panel) and $G^{TLT}(t_1,t_2)$
(right panel) at wave-vectors $k=k_0$ and $q=3k_0$ and time arguments
$t_1=3t$, $t_2=t$. In both panels, the un-connected circles are the
simulation data, the solid, dotted and dashed lines represent the full
MCT theory, the Euler MCT theory, and the Gaussian MCT theory
results, respectively.  The error estimates represent $96\%$
confidence intervals.

\vspace{1in}

\begin{table}[tpb]
\begin{tabular}{|r|c|c|}
&$TL;T$&$TT;L$\\\hline
$M^{21}_{\vec k-\vec q,\vec q; \vec k}$ & $i k\beta^{-1}$&  $i\sfrac23 kp/\rho$\\
$V^{21}_{\vec k-\vec q,\vec q; \vec k}$ & $i k\beta^{-1}$&  $-ik\beta^{-1}$
\end{tabular}
\caption{Expressions for the leading behavior of the {\em Euler}
vertices $\bar M^{21}$ and their Gaussian
counterparts $V^{21}$.}
\end{table}

\begin{table}[tpb]
\begin{tabular}{|r|c|c|c|}
&$TH;T$&$TN;T$&$TT;H$\\\hline
$M^{21}_{\vec k-\vec q,\vec q; \vec k}$ & 
$ - k(k-q) \left[ v_h - \sqrt{\frac23}\frac{\eta}{m\rho}\right]$
& 
$- k(k-q) v_n + k^{2} S(q) \frac{\eta}{m\rho}$
& $-k^{2} \left( v_{th}+\frac{4m}{3 \sqrt{6} \beta}
\frac{\lambda}{k_{B}} \right)$\\
$V^{21}_{\vec k-\vec q,\vec q; \vec k}$ & 
$ - k(k-q)  v_h - kq\frac{\eta}{m\rho}$
& $- k(k-q) v_n$ & $-k^{2} v_{th}$\\
\end{tabular}
\caption{Expressions for leading behavior of the {\em dissipative} 
vertices $\bar M^{21}$ and their Gaussian
counterparts $V^{21}$. Note that in the table $k$ and $q$ stand for
the $x$ component of $\vec k$ and $\vec q$, respectively.}
\end{table}


\begin{thebibliography}{99}    
\bibitem{Sillescu96}
  R. B\"{o}hmer, G. Hinze, G. Diezemann, B. Geil, and H. Sillescu,
Europhys. Lett. {\bf 36}, 55 (1996).
\bibitem{Spiess98}
  U. Tracht, M. Wilhelm, A. Heuer, H. Feng, K. Schmidt-Rohr, and
H.W. Spiess, Phys. Rev. Lett. {\bf 81}, 2727 (1998).
\bibitem{Fleming}
  D.A. Blank, L.J. Kaufman, and G.R. Fleming, J. Chem. Phys. {\bf
113}, 771 (2000).
\bibitem{Miller}
  V. Astinov, K.J. Kubarych, C.J. Milne, and R.J.D. Miller,
Chem. Phys. Lett. {\bf 327}, 334 (2000).
\bibitem{Tokmakoff}
  O. Golonzka, N. Demird\"{o}ven, M. Khalil, and A. Tokmakoff,
J. Chem. Phys. {\bf 113}, 9893 (2000).
\bibitem{Mukamel99}
  S. Mukamel, A. Piryatinski, and V. Chernyak, Acc. Chem. Res. {\bf
32}, 145 (1999).
\bibitem{Steffen97}
  T. Steffen, and K. Duppen, J. Chem. Phys. {\bf 106}, 3854 (1997);
Phys. Rev. Lett. {\bf 76}, 1224 (1996).
\bibitem{Keyes2000}
  T. Keyes and J.T. Fourkas, J. Chem. Phys. {\bf 112}, 287 (2000).

\bibitem{Kob}
  W. Kob, C. Donati, S.J. Plimpton, P.H. Poole, S.C. Glotzer,
Phys. Rev. Lett. {\bf 79}. 2827 (1997).

\bibitem{Glotzer}
  S. C. Glotzer, V. N. Novikov, and T. B. Schr\o der,
J. Chem. Phys. {\bf 112}, 509 (2000).
\bibitem{Heueretal}
  A.~Heuer and K.~Okun, J.\ Chem.\ Phys. {\bf 106}, 6176 (1997);
  A.~Heuer, Phys.~Rev.\ {\bf E56}, 730 (1997);
  B.~Doliwa and A.~Heuer, Phys.~Rev.\ Lett.\ {\bf 80}, 4915 (1998);
  B.~Doliwa and A.~Heuer, J.\ Phys.: Condens. Matter {\bf 11}, A277
(1999).

\bibitem{Tanimura97}
  K. Okumura and Y. Tanimura, J. Chem. Phys. {\bf 106}, 1687 (1997).

\bibitem{Stratt}
  A. Ma and R.M. Stratt, Phys. Rev. Lett. {\bf 85}, 1004 (2000).
\bibitem{Reichman}
  R.A. Denny and D.R. Reichman, Phys. Rev. E. {\bf 63}, 065101(R) (2001).
\bibitem{vanZon}
  R. van Zon and J. Schofield, {\it submitted} 

\bibitem{MO82}
  J.~Machta and I.~Oppenheim, Physica {\bf 112 A},  361  (1982).

\bibitem{SLO92}
  J.~Schofield, R.~Lim, and I.~Oppenheim, Physica {\bf 181A},  89
(1992).

\bibitem{K70}
	K.~Kawasaki, Ann. Phys. {\bf 61},  1  (1970).

\bibitem{R81}
	D.~Ronis, Physica {\bf 107A},  25  (1981).
\bibitem{Carnahan}
  N.F. Carnahan and K.E. Starling, J. Chem. Phys. {\bf 51}, 635
(1969).
\bibitem{Chapman}
  	S. Chapman and T.G. Cowling, {\it The Mathematical Theory of Non-Uniform Gases}
 	(Cambridge University Press, Cambridge, 1970).

\bibitem{eventMD}
  B.J. Alder and T.E. Wainwright, J. Chem. Phys. {\bf 27}, 1208 (1957).

 \bibitem{Zwanzig61}
	R. Zwanzig, in {\it{Lectures in Theoretical Physics}}
	(W.E. Britton, B. W. Downs and J. Downs eds., Wiley, New
York,1961); 
	R. Zwanzig, Ann. Rev. Phys. Chem. 
	{\bf	16}, 667 (1965).
\bibitem{Mori65}
	H. Mori, Prog. Theor. Phys. {\bf 33}, 423 (1965); H. Mori,
	Prog. Theor. Phys. {\bf 34}, 399 (1965).



\bibitem{BerneandPecora}
	B.J. Berne and R. Pecora, {\it Dynamic Light Scattering},
(John Wiley \& Sons, New York, 1976).
\bibitem{Gotze75}
        W. G\"{o}tze and M. L\"{u}cke, Phys. Rev. A {\bf 11}, 2173
        (1975).

\bibitem{Bosse78}
        J. Bosse, W. G\"{o}tze, and M. L\"{u}cke, Phys. Rev. A {\bf
        17}, 434 (1978); {\it Ibid}, 447 (1978); {\it Ibid}, Phys. Rev. A {\bf
        18}, 1176 (1978).  
\bibitem{dSC80}
	I.M. de Schepper and E.G.D. Cohen, Phys. Rev. A {\bf  22}, 287
(1980); J. Stat. Phys. {\bf 27}, 223 (1982).
\bibitem{Kirkpatrick85}
	T.R. Kirkpatrick, Phys. Rev. A {\bf 32}, 3130 (1985).

\bibitem{R80}
        D.C. Rapaport, J. Comp. Phys. {\bf 34}, 184 (1980).
  
\bibitem{KCdS87}
        B. Kamgar-Parsi, E.G.D. Cohen, I.M. de Schepper,
        Phys.\ Rev.\ A {\bf 35}, 4781 (1987).  

\bibitem{timeseries}
	M.B. Priestly, {\it Spectral Analysis and Time Series} (Wiley,
New York, 1971).
\bibitem{LimThesis}
	R. Lim, Ph.D. thesis, M.I.T. (1985).

\end{thebibliography}
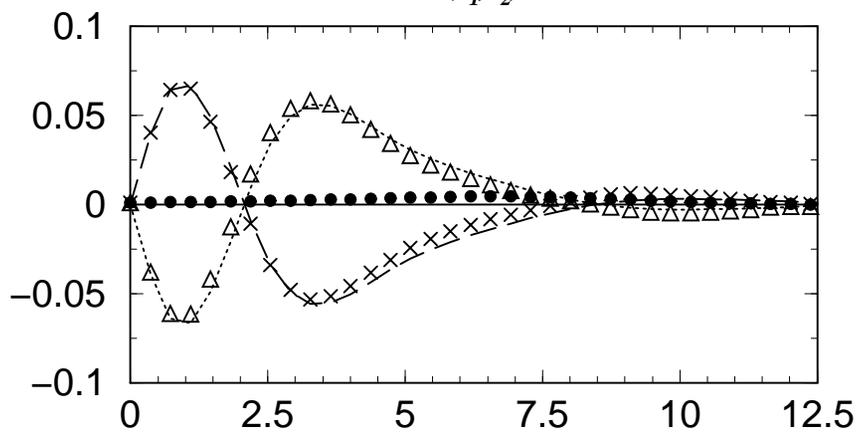
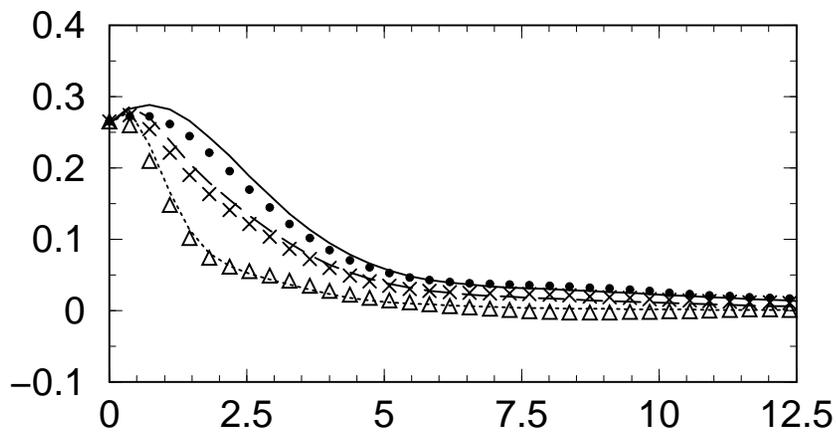
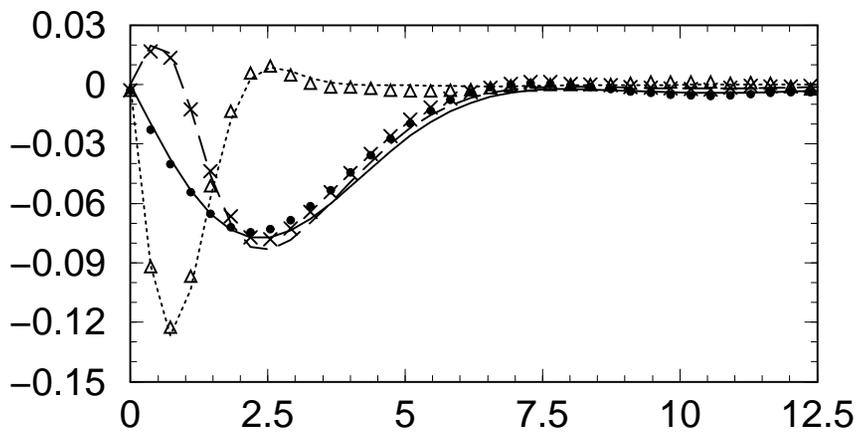
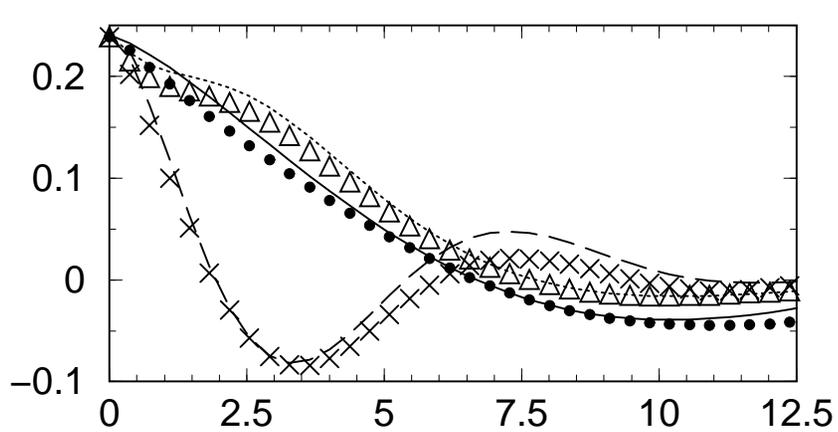
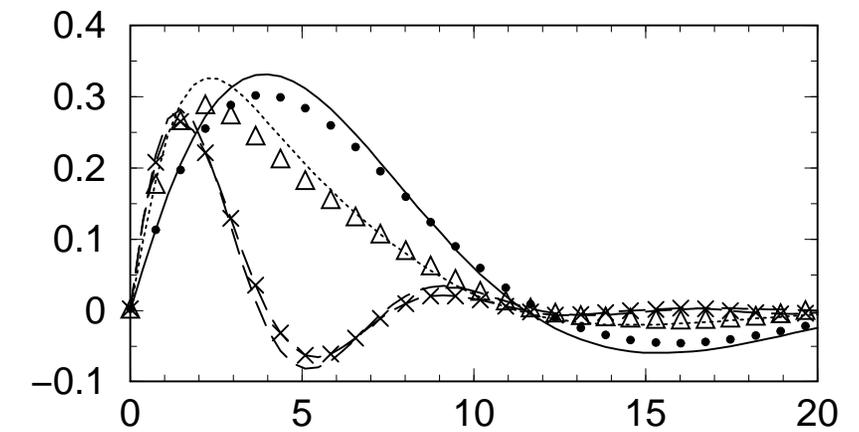
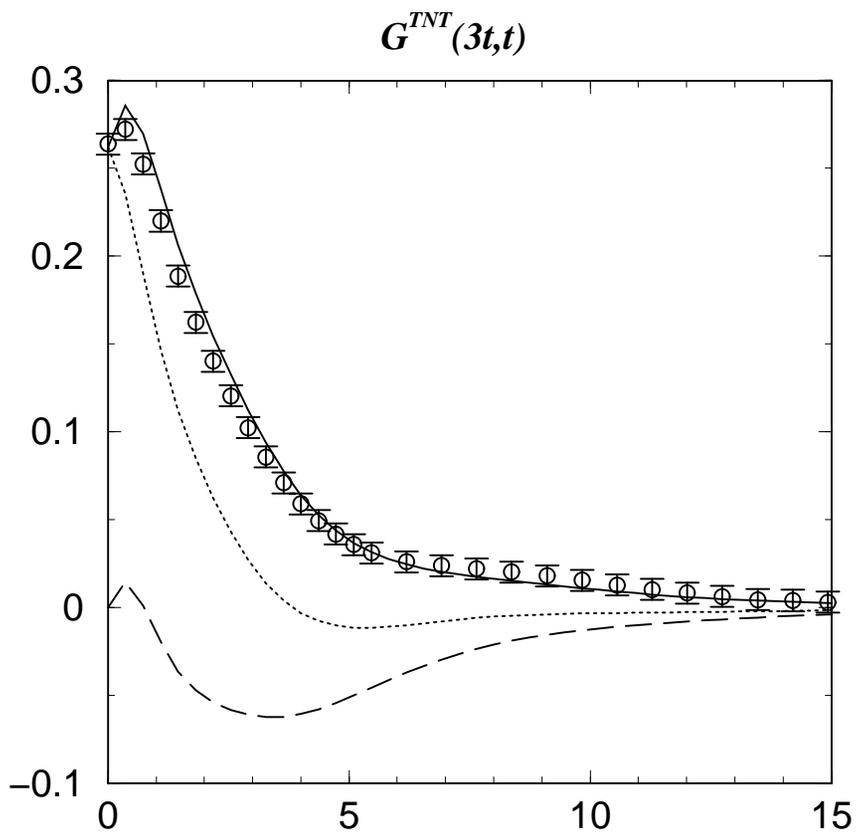
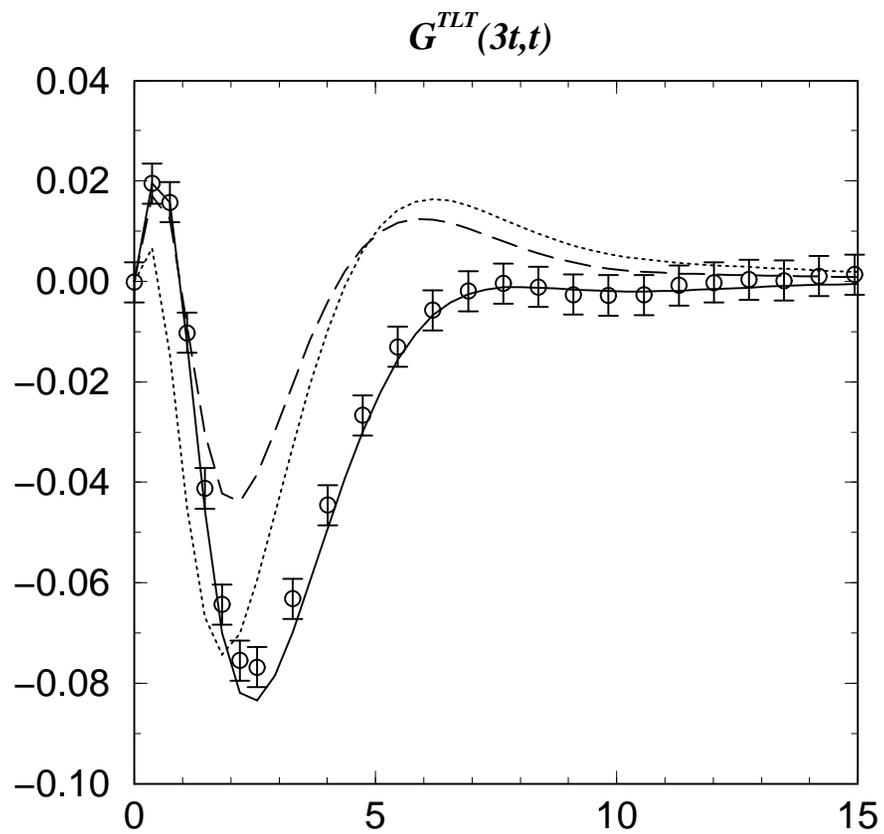
\end{document}